\documentclass[floatfix,twocolumn]{revtex4}

\usepackage{graphicx}

\begin{document}

\title{Optical properties of silicon nanoparticles in the presence of water:
       A first principles theoretical analysis}
\author{David Prendergast, Jeffrey C. Grossman, Andrew J. Williamson,
        Jean-Luc Fattebert and Giulia Galli} 
\affiliation{Lawrence Livermore National Laboratory, L-415, P.O.  Box 808, 
             Livermore, CA 94551.}

\date{\today}


\begin{abstract}
We investigate the impact of water, a polar solvent, 
on the optical absorption of
prototypical silicon clusters with oxygen passivation.
We approach this complex problem by assessing the contributions of
three factors: chemical reactivity; thermal equilibration and dielectric
screening. We find that the silanone (Si=O) functional group is not
chemically stable in the presence of water and exclude this as a source
of significant red shift in absorption in aqueous environments.
We perform first principles molecular dynamics simulations of
the solvation of an oxygenated silicon cluster with explicit water molecules
at 300~K. We find a systematic 0.7~eV red shift in the absorption gap of this
cluster, which we attribute to thermal strain of the molecular structure.
Surprisingly, we find no observable screening impact of the solvent, in
contrast with consistent blue shifts observed for similarly sized
organic molecules in polar solvents.
The predicted red shift is expected to be significantly smaller
for larger Si quantum dots produced experimentally, guaranteeing that their
vacuum optical properties are preserved even in aqueous environments.
\end{abstract}

\pacs{To be determined}

\maketitle

\section{Introduction}
\label{Sec.Introduction}

Molecular sensing technology currently relies on fluorescent
organic molecules as optical tags. 
However, these organic dyes have some deficiencies.
They absorb light only at specific frequencies, 
making them inefficient emitters.
Their optical emission cannot be tuned readily, and
many molecules with unrelated structures and chemistry
are required to produce a range of emission wavelengths.
Furthermore, prolonged irradiation can lead to chemical or structural
decomposition, altering or even destroying desirable
optical properties during experiments.
These deficiencies may be overcome by using semiconductor quantum dots (QDs)
to develop new optical sensing technology.

QDs are pieces of bulk semiconductor in the nanometer-size regime.
Their small size bestows them with optical properties that lie somewhere between
those of molecules and those of the corresponding bulk materials.
Like their bulk counterparts, QDs have robust atomic structures.
They absorb light efficiently over a wide range of wavelengths,
while emitting light of a characteristic frequency.
In addition, due to quantum confinement effects, 
the optical properties of QDs may
be tuned by controlling the size of each nanoparticle.
These useful properties make QDs highly competitive with organic dyes
for use as optical tags.

QDs made of silicon (Si) are highly desirable, given the abundance of
this element and its use in the electronics industry.
Also, in contrast with existing QD materials such as cadmium selenide (CdSe),
porous silicon (one of the precursors of Si QDs) 
is both bio-stable and nontoxic.~\cite{MayneBayliss2000}
Currently, the use of CdSe QDs as optical tags in biological environments
requires coating with many layers of materials,
to increase bio-stability, and shield the
biological environment from their inherent toxicity.
Si QDs do not require any coating and therefore
the overall size of Si QDs
is typically much smaller than that of coated CdSe QDs.
This should permit the use of Si QDs, 
which possess desirable optical properties 
and may be applied in aqueous environments,~\cite{HarwellCroney2003}
to probe biochemical processes currently
inaccessible to CdSe QDs -- 
specifically processes which involve diffusion through 
intra-cellular membranes.

However, at present, Si QDs of a particular size, 
synthesized using different approaches, 
have a large distribution of measured optical properties, 
e.g. photoluminescence from UV to red.~\cite{WilcoxonSamara1999} 
To shed light on the discrepancies between different experiments, 
first principles calculations have illustrated 
the important role of surface 
chemistry~\cite{FilonovOssicini2002,VasilievChelikowsky2002,
PuzderWilliamsonPRL2002,
PuzderWilliamsonJCP2002,PuzderWilliamsonJACS2003,
ZhouBrus2003,ZhouFriesner2003}
and structure~\cite{WeisskerFurthmuller2003,VasilievMartin2002,
PuzderWilliamsonPRL2003,DraegerGrossmanPRL2003,DraegerGrossmanJCP2004}
on the optical properties of Si QDs.
From this body of work, it is now clear that the optical properties 
of Si QDs are not determined by size alone.

In this work we analyze the impact of
water on the optical absorption properties of Si nanoparticles.
In the synthesis process or the application environment, 
water may be present, in vapor or liquid form,
as a solvent or a contaminant.
Furthermore, in applications to sensing in biological environments,
water would be ubiquitous.
Given the known impacts of water -- a polar solvent -- 
on the absorption properties of solvated
organic molecules, and the absence of equivalent information 
for inorganic solutes, 
analysis for Si nanoparticles in water is clearly necessary
and important.

\section{Consequences of solute-solvent interaction for optical properties}
\label{Sec.SoluteSolventInteractions}

In our study, we consider the solvation of an optically-active 
species in a polar solvent.
The strength of the solute-solvent interaction determines how much 
the optical properties of the total solvated system deviate from those
of its constituents: the isolated solute and the solvent. 
If the electronic states involved in optical processes can be clearly associated
with the solute (rather than with the solvent) then we may speak of
the impact of the solvent on the solute, and that is the focus of this
paper.
\footnote{
  This picture is analogous to optically-active point defects in
  solids, where defect states are introduced
  within the optical band gap of the solid,
  such that the optical properties of the combined system
  can be attributed to the defect rather than to the solid. 
}
We consider three solute-solvent interactions contributing to changes 
in the optical properties of the solute: 
chemical reactivity; thermal equilibration; and
dielectric screening. 

Dielectric screening refers to the impact on the molecular and electronic
structure of the solute due to screening of Coulomb interactions 
caused by the finite electrical susceptibilities of the solvent and solute.
From the electrostatic point of view, this is a self-consistent phenomenon, 
where the charge density of the solute
establishes a polarization of the solvent, which in turn impacts the
solute charge density until some equilibrium is reached.
Concomitantly, long-range electron correlation leads to attractive
dispersion forces between the solute and solvent charge densities.
The impact of these screening contributions on optical properties
is explained in detail in Sec.~\ref{Sec.SolventScreening} and
the impact of screening on the Si clusters
studied here is discussed in Sec.~\ref{Sec.Screening}.
Sections~\ref{Sec.AbinitioCalculations} and~\ref{Sec.ApplicationSi5}
contain details of our first principles calculations and their application
to Si clusters {\it in vacuo}, respectively.

An investigation of the reactivity of various Si clusters with water is
given in Sec.~\ref{Sec.Reactivity}.
Clearly, if the solute is chemically reactive in the presence of the solvent,
the resulting reaction product may have a different molecular structure
or composition, and, consequently, a completely different optical signature.

Finally, in Sec.~\ref{Sec.gaps300K} we examine the effect of strain induced
by thermal fluctuations on optical properties of Si clusters.
Thermal equilibration with a reservoir of solvent molecules
will activate thermally accessible vibrational modes of the solute. 
Already, for Si clusters {\it in vacuo} at zero temperature, 
there is clear evidence that strain in the molecular structure, 
e.g. due to surface relaxation,
has a large impact on
the optical properties of the system.~\cite{PuzderWilliamsonPRL2003,
DraegerGrossmanPRL2003,WeisskerFurthmuller2003}

\section{Solvent screening and optical absorption}
\label{Sec.SolventScreening}

\begin{figure}[t]
  \begin{centering}
    \scalebox{0.6}{\includegraphics{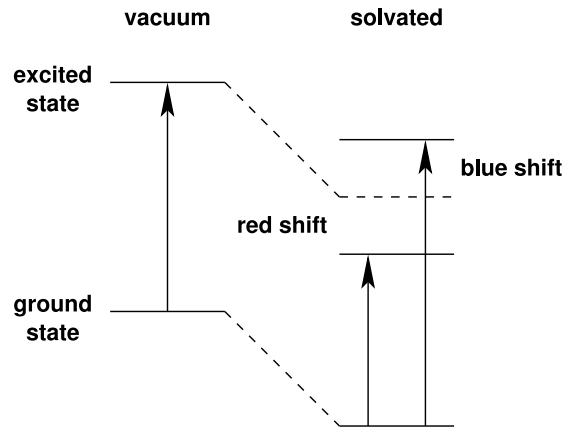}}
  \end{centering}
  \caption{
    Schematic energy level diagrams for the electronic states 
    involved in optical absorption for vacuum and solvated systems.
    Absorption energies required for transitions from ground to
    excited states are indicated with arrows.
    The solvation energy causes shifts in the total energy of 
    the ground and excited states. 
    The difference in solvation energies of the ground and excited
    states leads to a red or blue shift in the absorption energy
    with respect to the absorption energy {\it in vacuo}.
  }
  \label{fig.solvshift}
\end{figure}

In order to study the impact of water on the optical absorption
of Si nanoparticles, it suffices to assess 
the impact of solute-solvent interactions on the electronic ground state 
and on the first excited state.
The solvation energy
\footnote{The solvation energy is defined as 
          $[E(\mbox{solute}) + E(\mbox{solvent})] 
           - E(\mbox{solute}+\mbox{solvent})$,
          where $E(\mbox{solute})$ is the total energy of the isolated solute;
          $E(\mbox{solvent})$ is the total energy of the pure solvent;
          and $E(\mbox{solute}+\mbox{solvent})$ is the total energy of the
          solvated system. 
}
may be different for ground and excited states, and this is the origin of
so-called solvation shifts (Fig.~\ref{fig.solvshift}).
If the solvation energy of the excited state is larger (smaller) 
than that of the ground state, we see a red (blue) shift in the 
absorption energy when compared with the unsolvated system.

One source of this difference in solvation energies is dielectric screening. 
For a large class of systems, probability distributions of excited state 
wave functions are more diffuse than those of the ground state, 
and consequently the system has a higher polarizability
in the excited state than in the ground state.
In general, the more polarizable a system, the larger its solvation energy.
Therefore, if the excited state is screened by the solvent, this
larger solvation energy will lead to a red shift in absorption.
This is often the case for non-polar solvents. 

In polar solvents a significant contribution to the screening 
comes from the alignment of molecular dipoles
and occurs on the timescale of molecular motion.
Upon absorption of a photon, there is not sufficient time for this
orientational polarization to screen the resulting change in the
electronic charge density of the solute. Hence, the resulting
{\it non-equilibrium} excited state has a reduced solvation energy with
respect to the ground state, leading to an overall blue shift in
absorption (Fig.~\ref{fig.solvshift}).

In practice, both the fast and slow components of the screening response
of the solvent are present, leading to a net solvent shift.
Water is a highly polar solvent, with an average dipole moment per molecule
of $\sim1.9$~Debye in the gas phase. 
The electrical susceptibility of water can be separated
into a static component $\chi_0$ and an instantaneous component $\chi_\infty$.
($\chi_0 = \epsilon_0 - 1$, where $\epsilon_0$ is the static dielectric 
constant, and $\chi_\infty = \epsilon_\infty -1 = n^2 -1$, where
$\epsilon_\infty$ is the optical dielectric constant and $n$ is the
refractive index).
For liquid water at room temperature, $\chi_\infty = 0.78$ and $\chi_0 \sim 77$.
Experimental estimates of the Debye relaxation time for water provide
an estimate of the time taken for reorganization of the
orientational polarization in response to an external electric field.
Current theories and experiments agree that there are
two Debye relaxation times for water, one fast and one slow. The fast
relaxation time has been tentatively associated with single molecular
reorientation and has values from 0.2 -- 1.5~ps 
at ambient conditions.~\cite{Gaiduk2003,BuchnerBarthel1999} 
Therefore we can assume that electronic absorption processes
(occuring on 0.1~fs time scales) are instantaneous and that the
excited states correspond to a non-equilibrium nuclear configuration.
In addition, the large difference in magnitudes between $\chi_\infty$ and
$\chi_0$ increases the energetic cost of
the excited state and generally leads to blue shifts in the 
absorption spectra of
optically-active polar organic molecules, 
e.g. acetone (0.2~eV)~\cite{BernasconiSprik2003} and
pyridazine (0.5~eV).~\cite{BabaGoodman1966}
In this paper, we will see (surprisingly) 
that these characteristic blue shifts are absent
(or at least negligible in the presence of thermal strain) 
for small inorganic Si clusters
(see Sec.~\ref{Sec.Screening}).

\section{First principles calculations}
\label{Sec.AbinitioCalculations}

\begin{figure}[t]
  \begin{centering}
    \scalebox{0.30}{\includegraphics{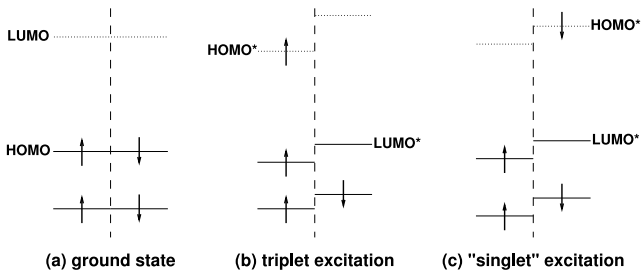}}
  \end{centering}
  \caption{
    Schematic energy level diagrams of the electronic states 
    considered in this paper, where single particle orbitals are indicated
    by horizontal lines, energy increases in the vertical direction,
    and up and down spin channels are separated with a dashed line.
    (a) The spin degenerate ground state, indicating the
    highest occupied molecular orbital (HOMO) and the lowest unoccupied 
    molecular orbital (LUMO). 
    (b) The first excited state with triplet 
    symmetry in a single reference picture, where the highest occupied
    (lowest unoccupied) molecular orbital of this excited state configuration,
    labeled HOMO$^*$ (LUMO$^*$) can be associated with the LUMO (HOMO)
    of the ground state. 
    (c) the first excited state used to approximate singlet symmetry,
    with the HOMO$^*$ and LUMO$^*$ in the same spin channel.
  }
  \label{fig.elevels}
\end{figure}

The optical properties of the various Si clusters examined in this paper 
are known to be particularly sensitive to their molecular 
structure.~\cite{PuzderWilliamsonPRL2003}
Furthermore, the properties of the solvent and the nature of the
solute-solvent interaction can be quite complex, involving charge-transfer,
hydrogen-bonding and long-range screening.
To capture all of this in our theoretical approach, we need highly
accurate estimation of the electronic structure of these systems.
For these reasons we adopt first principles computational techniques.

We make use of density functional theory (DFT) under the
generalized gradient approximation (GGA).
To simulate the impact of finite temperature and explicit solute-solvent
molecular interactions, we use first principles molecular dynamics (FPMD),
employing the Car-Parrinello method,\cite{CarParrinello1985}
treating the nuclei in the system as classical
particles acted upon by Coulombic inter-ionic forces and quantum forces
derived from the electronic structure.

The majority of electronic structure calculations in this paper 
were performed using the
plane-wave pseudo-potential codes {\sc gp}~\cite{GP} and 
{\sc abinit}.~\cite{ABINIT} 
We also used
the all electron code {\sc gaussian 98}.~\cite{Gaussian98} 
FPMD simulations were carried out using {\sc gp}. 
The DFT/GGA plane-wave pseudo-potential calculations make use of the PBE
exchange correlation functional~\cite{PerdewBurke1996} and
use a basis of planewaves consistent with periodic boundary conditions on
a simulation box of a given size. All calculations are performed for a
single point in the Brillouin zone, namely, the $\Gamma$-point.
In these plane-wave calculations, 
only valence electrons are included, with atomic
cores represented by nonlocal, 
norm-conserving pseudo-potentials.~\cite{Hamann1989,TroullierMartins1991} 
The Gaussian-basis all electron calculations make use of 
the Becke functional~\cite{Becke1988}
for exchange and the PW91 correlation functional.~\cite{WangPerdew1991} 
These calculations are performed for finite systems using a 
6-311G** Gaussian basis set.

A plane-wave kinetic energy cut-off of ~70 Ry
\footnote{
  In practice, we use a plane-wave cut-off of 69 Ry, which leads
  to more efficient grid dimensions for fast Fourier transforms.
}
produces converged results
for the electronic structure of the oxygen-containing Si
clusters considered here and for water in the condensed phase. 
For calculations involving Si clusters {\it in vacuo},
simulation cell dimensions are chosen appropriately to reduce the
impact of finite size effects on the results. This is particularly
important in calculations on small molecules where the spatial 
extent of unoccupied
electronic states can be much larger than the atomic dimensions of the
molecule. 

As discussed in Sec.~\ref{Sec.SolventScreening}, to determine the
absorption properties of a given system, we require explicit knowledge
of the equilibrium electronic ground state and the non-equilibrium
excited state. In practice, the absorption spectrum of a physical system
is composed of many allowed electronic transitions and associated vibrational
transitions. In this paper, we concentrate only on the lowest possible
electronic transition in the absence of vibrational excitations.
This marks the onset of optical absorption and is equivalent to the
band gap of a solid state system. 
We shall frequently refer to this energy as the absorption gap, $E_g$.
In assessing the impact of water on the optical properties of Si clusters,
we assume that a change in the absorption gap is indicative of an overall
shift in the absorption spectrum of these systems.

Estimation of the absorption gap using first principles methods can be
difficult. Electronic excitations are at heart many-body processes,
and so, are difficult to describe using effective single-particle DFT orbitals,
especially since most correlation functionals are designed specifically for
the electronic ground state.
One common DFT estimate of $E_g$ is the difference between the
Kohn-Sham eigenvalues of the highest occupied molecular orbital (HOMO)
and the lowest unoccupied molecular orbital (LUMO) of the ground
state electronic structure [Fig.~\ref{fig.elevels}~(a)].
For finite systems {\it in vacuo}, This HOMO-LUMO gap has been shown,
to follow the same trends, as a function of size, as more
accurate estimates of the absorption gap computed using 
quantum Monte Carlo (QMC) calculations.~\cite{WilliamsonGrossman2002}

However, in determining the impact of dielectric screening on the
optical properties of a solvated system, 
we note that the DFT HOMO-LUMO gap is not an
appropriate estimator of the absorption gap. 
Since the LUMO is never occupied, the HOMO-LUMO gap does not include
physical information on the change in solute charge density and
corresponding response in solvent polarization 
upon electronic excitation.
For solvated systems, 
we choose to employ the $\Delta\mbox{SCF}$ approach, to determine
the absorption gap. Here, we define $E_g = E^*_{tot}(X_0) - E_{tot}(X_0)$, where
$E^*_{tot}(X_0)$ is the total excited state energy and $E_{tot}(X_0)$ is the
total ground state energy, both evaluated for the ground state molecular
structure $X_0$. By construction, $E^*_{tot}$ includes the 
impact of electron density
variation associated with the solute excited state and any screening impacts
from the solvent in response to the solute.
The importance of the $\Delta\mbox{SCF}$ estimator of $E_g$ 
for solvated systems is discussed in Sec.~\ref{Sec.ElectronicStructure}.

For the systems considered here, the first allowed electronic transition 
from the ground state is to a state with singlet symmetry. 
The representation of this singlet state 
requires more than one reference electron configuration.
Since DFT is an effective single particle method we approximate
this excitation with the single reference state
described in Fig.~\ref{fig.elevels}(c).
This approximate ``singlet'' does not have the same symmetry as the 
true singlet,
but it does contain a realistic Coulomb interaction between the
states near the Fermi level.
Furthermore, a QMC calculation of the absorption gap of 
the silane molecule (SiH$_4$)
using this single reference approximation has been shown 
to be accurate to within 0.1~eV.~\cite{GrossmanRohlfing2001} 

However, it is well known that numerical convergence of the DFT 
``singlet'' first excited state energy is inherently difficult.
A more practicable energy to calculate is that of the
first excited electronic state with triplet symmetry 
[Fig~\ref{fig.elevels}~(b)].
This triplet state has a reduced Coulomb energy when compared with the
``singlet'', due to the exchange repulsion that keeps like-spin 
electrons apart.
For reasons of efficiency and robustness, we choose to compute the
$\Delta\mbox{SCF}$ estimate of $E_g$
using the triplet excited state. In practice, using either the
``singlet'' or triplet excited state energy 
does not qualitatively change our results,
and we provide evidence for this in 
Sec.~\ref{Sec.ElectronicStructure} by comparing with some ``singlet''
state calculations.
In the rest of this paper, we shall refer to the ``singlet'' state without
quotation marks.

\section{Silicon clusters in the absence of water}
\label{Sec.ApplicationSi5}

\begin{figure}[t]
  \begin{centering}
    \scalebox{0.55}{\includegraphics{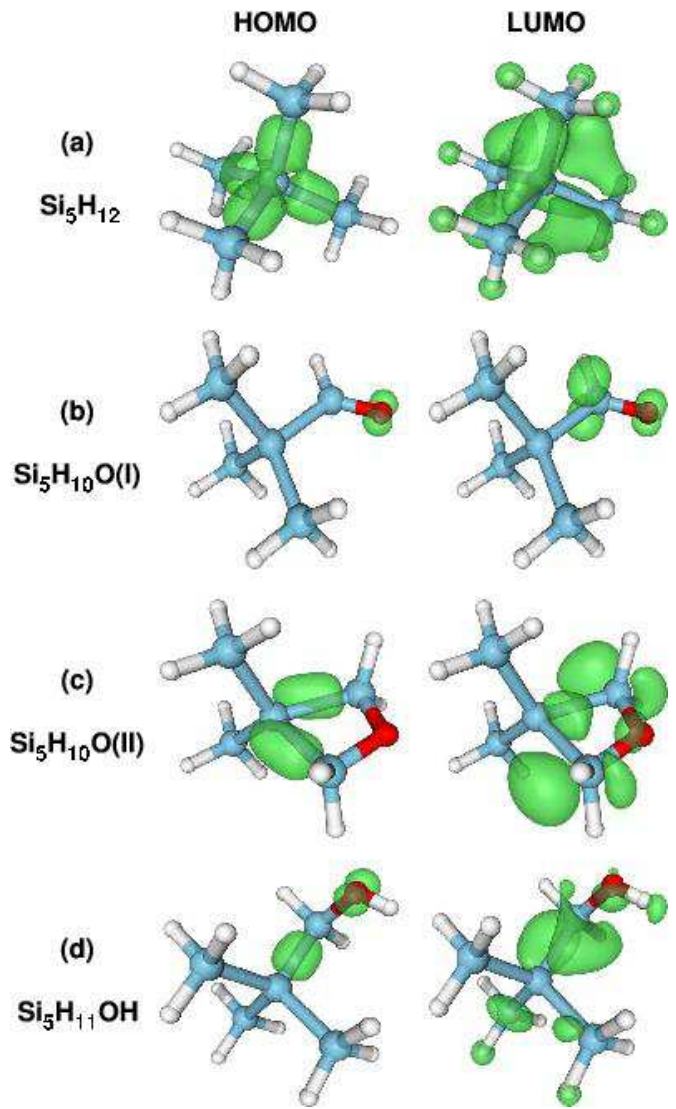}}
  \end{centering}
  \caption{
    The structures of various silicon clusters and corresponding 
    probability distributions (green) of the
    highest occupied molecular orbital (HOMO) and lowest unoccupied
    molecular orbital (LUMO).
    All electron density isosurfaces contain 30\% of the total charge 
    of the given state.
    Si atoms are colored blue, O red and H white.
    }
  \label{fig.homolumo}
\end{figure}

\begin{table}[t]
\caption{Absorption gaps (eV) and dipole moments (Debye) 
         for various silicon clusters. Gaps estimated using:
         HOMO-LUMO DFT/PBE pseudo-potential calculations (PBE-pp);
         HOMO-LUMO DFT/PW91 Gaussian basis set, all-electron calculations
         (PW91-ae);
         and ``singlet'' $\Delta\mbox{SCF}$ QMC calculations (QMC).
         Dipole moments estimated from PBE-pp charge densities ($d_{vac}$).
         Also shown are dipole moments calculated in the 
         presence of a continuum dielectric
         solvation model for water ($d_{solv}$) 
         and the angular change in direction ($\theta$) from the
         unsolvated dipole moment.
         (The symmetry of Si$_{5}$H$_{12}$ guarantees that its dipole
         moment is zero.)
        }
\begin{tabular*}{\columnwidth}{@{\extracolsep\fill}lccc@{\hspace{0.4cm}}ccc}
  \hline \hline
  cluster & \multicolumn{3}{c}{$E_g$} & \multicolumn{3}{c}{dipole} \\
     & PBE-pp & PW91-ae & QMC & $d_{vac}$ & $d_{solv}$ & $\theta$ \\
  \hline
  Si$_{5}$H$_{12}$             & 5.8 & 6.0 & 6.8 & 0.0 & 0.0 & $0.0^\circ$ \\
  Si$_{5}$H$_{10}$O (I)        & 2.9 & 3.1 & 3.6 & 4.5 & 6.2 & $8.5^\circ$ \\
  Si$_{5}$H$_{10}$O (II)       & 4.5 & 4.9 & 5.2 & 1.6 & 2.5 & $3.1^\circ$ \\
  Si$_{5}$H$_{11}$OH           & 4.7 & 5.0 & 5.4 & 1.2 & 2.0 & $4.9^\circ$ \\
  \hline \hline
\end{tabular*}
\label{tab.gaps}
\end{table}

The Si clusters studied in this paper are presented in Fig.~\ref{fig.homolumo}
and the corresponding estimates of the absorption gaps {\it in vacuo}
are presented in Table~\ref{tab.gaps}. 
Despite definite inadequacy in the DFT estimates of the absorption gap, when
compared with accurate QMC calculations, there are still consistent trends in
comparison of the gaps of $\mbox{Si}_5$ clusters with different passivants. 
This has also been demonstrated in previous work over a wide range of
Si cluster sizes.~\cite{PuzderWilliamsonPRL2002}
Also, the smaller systematic differences between the two sets of DFT gaps are
due to the choice of exchange correlation functional and whether the
pseudo-potential approximation is adopted.
We exploit the consistent trends in these calculated gaps, which
allow qualitative comparison of the gaps of different systems as
long as they are computed using the same level of theory.

Our reference structure is the ``pure''
hydrogen-saturated $\mbox{Si}_5\mbox{H}_{12}$ [Fig.~\ref{fig.homolumo}(a)].
We also consider various types of surface passivation containing oxygen.
The first, $\mbox{Si}_5\mbox{H}_{10}\mbox{O(I)}$ contains a silanone 
$(\mbox{Si=O})$ group [Fig.~\ref{fig.homolumo}(b)]. 
The departure from the local tetrahedral symmetry of each Si atom 
to a planar $sp^2$ symmetry for the SiHO portion of the molecule 
produces a localized HOMO and LUMO on this functional group. 
The localization effectively creates a
defect state within the ``band gap'' of the molecule reducing the
absorption gap by almost 50\%. We refer to previous work for a discussion of
this effect for Si clusters of various sizes.~\cite{PuzderWilliamsonPRL2002}

We next consider two other forms of oxygen passivation which maintain,
to a greater degree, the local tetrahedral symmetry of the Si atoms 
in the cluster. 
$\mbox{Si}_5\mbox{H}_{10}\mbox{O(II)}$ contains a bridging oxygen atom
between neighboring Si atoms at the surface
[Fig.~\ref{fig.homolumo}(c)]. This causes some strain
in the angle between the surface Si atoms bonded to the O and the core
Si atom, reducing the tetrahedral angle from the ideal value 
(109.5$^\circ$) to 70.4$^\circ$.
However, the preservation of approximate tetrahedral symmetry 
around each Si atom prevents the same degree of localization
of the HOMO and LUMO as we see in $\mbox{Si}_5\mbox{H}_{10}\mbox{O(I)}$.
Consequently, the impact on the absorption gap is diminished and,
for $\mbox{Si}_5\mbox{H}_{10}\mbox{O(II)}$, $E_g$ is 25\% less than
that of $\mbox{Si}_5\mbox{H}_{12}$. 
$\mbox{Si}_5\mbox{H}_{11}\mbox{OH}$ contains a hydroxyl (OH) passivant, 
which has a reduced impact on the molecular structure in terms of strain,
in comparison with $\mbox{Si}_5\mbox{H}_{10}\mbox{O(I)}$. 
However, there is still a reduction in the absorption gap due to 
the electro-negativity of the O atom, 
which attempts to localize the states near the Fermi level around it
[Fig.~\ref{fig.homolumo}(d)].

In a first approximation to the impact of water on the structure of
these Si$_5$ clusters, we use a continuum solvation
model~\cite{FattebertGygi2002,FattebertGygi2003} to simulate the
electrostatic impact of the solvent. We find almost no change in
the atomic structure of these molecules in the presence of this
solvation model. The most noticeable change is a 1\% increase in
the bond length of the silanone group in $\mbox{Si}_5\mbox{H}_{10}\mbox{O(I)}$.
However, an examination of the impact of this solvation model on
the dipole moments of each molecule indicates a consistent tendency
to increase the dipole moment (Table~\ref{tab.gaps}).
Given that the atomic structure remains unperturbed in the presence of
the solvent polarization, this indicates that the increase in the dipole
moment is caused by polarization of the electron charge density.

\section{Reactivity of oxygenated silicon clusters with water}
\label{Sec.Reactivity}

\begin{figure}[t]
  \begin{centering}
    \scalebox{0.37}{\includegraphics{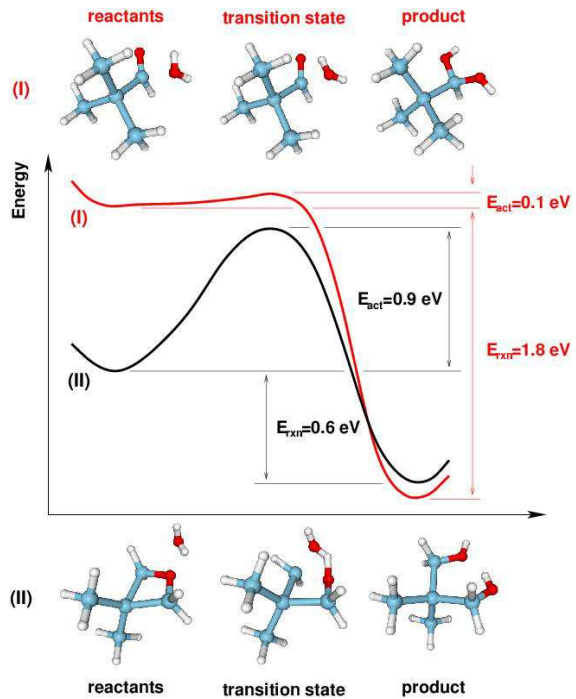}}
  \end{centering}
  \caption{
    The reaction paths of $\mbox{Si}_5\mbox{H}_{10}\mbox{O(I)}$ (red)
    and $\mbox{Si}_5\mbox{H}_{10}\mbox{O(II)}$ (black)
    with one water molecule, indicating the activation energy
    $E_{\mbox{act}}$ and the overall energy of reaction $E_{\mbox{rxn}}$.
    }
  \label{fig.reaction}
\end{figure}

\begin{table}[t]
\caption{Variations in the HOMO-LUMO gap (eV) during the reaction with one
         water molecule for $\mbox{Si}_5\mbox{H}_{10}\mbox{O(I)}$
         and (II). Calculated using PW91-ae.
        }
\begin{tabular*}{\columnwidth}{@{\extracolsep\fill}lcccccc}
  \hline \hline
  cluster     & reactants & transition state & product \\
  \hline
  Si$_{5}$H$_{10}$O (I)        & 3.8 & 4.0 & 5.3 \\
  Si$_{5}$H$_{10}$O (II)       & 4.7 & 4.1 & 4.7 \\
  \hline \hline
\end{tabular*}
\label{tab.reactiongaps}
\end{table}

The most striking impact of the solvent on a solute would be a chemical
reaction producing a more stable chemical species, which
we could expect to have quite different optical properties to its precursor.
Of the four clusters considered in Section \ref{Sec.ApplicationSi5},
$\mbox{Si}_5\mbox{H}_{10}\mbox{O(I)}$ readily reacts with water to produce a
dihydroxide according to the reaction:
\begin{equation}
  \mbox{Si}_5\mbox{H}_{10}\mbox{O(I)} + \mbox{H}_2\mbox{O}
  \Rightarrow \mbox{Si}_5\mbox{H}_{10}\mbox{(OH)}_2
\end{equation}

Using {\sc gaussian 98},
we determined a 0.1~eV activation energy
for this reaction. This involved finding the total energy of the
transition state for the reaction, by
structural optimization in the direction of the eigenvector of the
Hessian matrix with lowest negative eigenvalue.
This reaction profile is compared, in Fig.~\ref{fig.reaction},
with the same reaction for the stoichiometrically
equivalent bridging structure $\mbox{Si}_5\mbox{H}_{10}\mbox{O(II)}$,
which has a larger activation energy of 0.9 eV. 
This analysis indicates that the survival of silanone groups in water is
highly unlikely, given the comparable size of the activation energy
with the thermal energy at 300~K (i.e., $\sim0.03$~eV). Also, this reaction is
exothermic, and the energy released ($\sim1.8$eV) would
fuel the breakdown of other silanone groups by water.

Previous work by Zhou and Head~\cite{ZhouHead2000} indicates a larger 
activation energy of 0.3~eV for the silanone group's reaction with water. 
However, the level of theory (MP2) and smaller system size (one Si atom) 
in that work may lead to a larger activation energy than the one found
here. Furthermore, the
1\% increase in the Si=O bondlength in the presence of a polarizable
solvent model (Sec.~\ref{Sec.ApplicationSi5}) indicates that in liquid water 
this activation energy may be further reduced due to polarization of the 
$\mbox{Si}_5\mbox{H}_{10}\mbox{O(I)}$ molecule and weakening of the
Si=O bond.
A recent study of the reactivity of several silica clusters with water
by Laurence and Hillier~\cite{LaurenceHillier2003}, using the B3LYP
functional and a comparable basis to ours, reports activation energies of
$\sim0.9$~eV for the reaction of the Si-O-Si group with water,
in good agreement with our result for the reaction of 
$\mbox{Si}_5\mbox{H}_{10}\mbox{O(II)}$ with a water molecule. 

The reaction of $\mbox{Si}_5\mbox{H}_{10}\mbox{O(I)}$ with water leads
to a product with a larger absorption gap.
The restoration of the local $sp^3$ symmetry around the Si atom of
the Si=O group relaxes the strain on the electronic structure, and the
localized states present in the reactant disappear. This
increases the DFT HOMO-LUMO gap from 3.8~eV 
(in the presence of a water molecule)
to 5.3~eV, comparable to that of the single hydroxide.

We have also observed this reaction to take place in the course of a
FPMD simulation of $\mbox{Si}_5\mbox{H}_{10}\mbox{O(I)}$ in a simulation cell
containing 57 water molecules. This reaction occurred while the
system temperature was being raised to 300~K, within the first 0.2~ps
of the simulation.
Therefore, we exclude this cluster from further tests on absorption 
in the presence of water.
Given that the product of this reaction is a hydroxide, which is significantly
more stable than the silanone, we regard the hydroxide cluster as
a chemically stable species in water. 

The absorption gap of $\mbox{Si}_5\mbox{H}_{10}\mbox{O(II)}$ remains
essentially the same at 4.7~eV, following the reaction with water.
We regard this bridged cluster as chemically stable 
given its larger activation energy (0.9~eV),
and smaller exothermic energy (0.6~eV) 
in comparison with that of the silanone cluster. 
Furthermore, as we see in Sec.~\ref{Sec.gaps300K}, 
during a 4.5~ps FPMD simulation we do not observe
the hydroxylation of this molecule when solvated in liquid water, 
despite large fluctuations in the
cluster's average kinetic energy which were equivalent to temperatures
exceeding 1000~K [Fig.~\ref{fig.tempgap}(a)].

\section{Impact of finite temperature on optical properties of
            silicon clusters}
\label{Sec.gaps300K}

\begin{figure}[t]
  \begin{centering}
    \scalebox{0.45}{\includegraphics{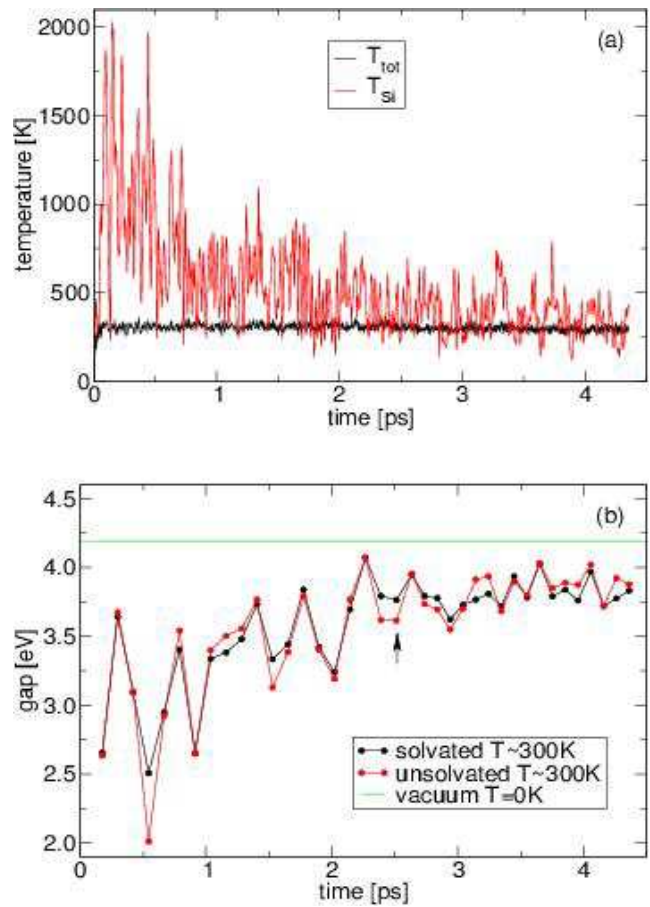}}
  \end{centering}
  \caption{
    (a) The system temperature $T_{tot}$ (black) and the instantaneous
    average kinetic energy (converted to units of temperature) of the
    Si atoms $T_{Si}$ as a function of time during an FPMD simulation of 
    $\mbox{Si}_5\mbox{H}_{10}\mbox{O(II)}$ in water.
    (b) The absorption gap of the solvated system (Si cluster and
    water molecules) computed
    at the indicated times during the course of the FPMD simulation (black)
    and the corresponding gap of the same Si cluster structure without the 
    water molecules (red). Also shown is the gap of the 
    Si cluster in its relaxed, zero temperature structure (green).
    The arrow indicates the particular snapshot used to produce the electronic
    structure information in Figs.~\ref{fig.states} and ~\ref{fig.dos}
    and Table~\ref{tab.gapcomparison}.
    }
  \label{fig.tempgap}
\end{figure}

We simulate the solvation of the molecule
$\mbox{Si}_5\mbox{H}_{10}\mbox{O(II)}$ in water using FPMD.
The simulation cell dimensions are 
$(12.169~\mbox{\AA} \times 12.169~\mbox{\AA} \times 12.169~\mbox{\AA})$ 
and the cell contains 57 water molecules in addition to the solute.
These cell dimensions were optimized to produce ambient pressure conditions 
using a classical MD simulation. The final configuration of this classical
simulation was used as the starting point of the FPMD run.

The initial 0.17~ps of the FPMD simulation evolve under the influence of a
velocity rescaling thermostat after which the temperature of the
system reaches 300~K. At this point the thermostat is removed and the
simulation is allowed to evolve with constant total energy.
\footnote{ 
  For the final 1.5~ps of the simulation the time step is increased to 3~a.u,
  while the fictitious electron mass was kept fixed at 276~a.u. 
}

We notice that the molecular structure
of the solute is greatly perturbed during this simulation.
This perturbation is caused by the non-equilibrium initial configuration
that is used to start the simulation. We assume that the classical
forces used to generate this initial configuration have some mismatch
with the quantum mechanical forces present in the FPMD simulation.
This mismatch greatly increases the average
kinetic energy of the solute, as can be seen in 
Fig.~\ref{fig.tempgap}(a) (the average kinetic energy of the Si atoms
in the solute, $T_{Si}$, is shown here converted to temperature
units for comparison with the system temperature $T_{tot}$). 
We see that the solute is effectively equilibrated with 
the thermal bath of water molecules after 2~ps.

For such a small molecule, it is reasonable to predict that the impact
of thermal fluctuations on the structure of 
$\mbox{Si}_5\mbox{H}_{10}\mbox{O(II)}$ may have a marked impact on
its optical properties. From previous work,~\cite{PuzderWilliamsonPRL2003,
DraegerGrossmanPRL2003,WeisskerFurthmuller2003}
we know that the impact of strain on Si clusters leads to large
red shifts in the absorption gap. This is also observed in our
finite temperature simulation. Taking snapshot structural configurations
approximately every 0.1~ps from this simulation, we approximate the
instantaneous absorption gap as the difference in total energy between
the ground state and the triplet first excited state. 
This is shown in the black curve of Fig.~\ref{fig.tempgap}(b).
It is clear that variations in
the absorption gap are highly correlated with the average kinetic energy
(or instantaneous ``temperature'') of the Si atoms,
shown in Fig.~\ref{fig.tempgap}(a). 
The larger fluctuations in the absorption gap occur at 
higher temperatures, in the
2~ps equilibration phase of the simulation, where the corresponding structural
fluctuations are largest, producing the greatest strain.
Once equilibrated, the fluctuations in the gap reach a steady
standard deviation of $\sim0.3$~eV. 
The equilibrated system displays a systematic red shift in the absorption
gap of 0.7~eV from that of the solute {\it in vacuo} at 0~K.

\section{Impact of screening on optical absorption of silicon clusters}
\label{Sec.Screening}

The observed 0.7~eV shift is not an effect due to polarization or
dispersion from the surrounding solvent molecules. To verify this, we
remove the water molecules from each structural configuration used 
previously and compute the absorption gap of the isolated 
$\mbox{Si}_5\mbox{H}_{10}\mbox{O(II)}$ molecule.  
We find that the resulting unsolvated gaps [red curve in 
Fig.~\ref{fig.tempgap}(a)] follow the corresponding solvated
gaps quite closely. We observe no systematic shift to the red or blue
due to the presence of the solvent in the equilibrated phase of the
simulation. This is a clear indication that the dominant factor influencing
the absorption properties of the solute is the finite temperature of
the system.

It is also interesting to note that even in the highly non-equilibrium phase
of the simulation of the first 2~ps, structural variations remain
the dominant factor in variations of the absorption gap. 
In this regime, the solvated and unsolvated gaps are practically the
same, while fluctuating over a range of 1.5~eV.

The absence of a systematic solvent shift, due to screening, for the
Si cluster in water is at variance with the solvent shifts of polar
organic solutes in polar solvents, where an experimentally observed
blue shift in the absorption gap is expected from screening arguments.
Recent calculations by 
Bernasconi {\it et al.}\cite{BernasconiSprik2003}
have shown that a solvent blue shift for 
acetone in water is resolvable from a
relatively short FPMD simulation.
This contrast with our results is interesting and may point 
to fundamental differences
between organic and inorganic solutes in water.

\section{Electronic Structure}
\label{Sec.ElectronicStructure}

\begin{figure*}[t]
  \begin{centering}
    \scalebox{0.7}{\includegraphics{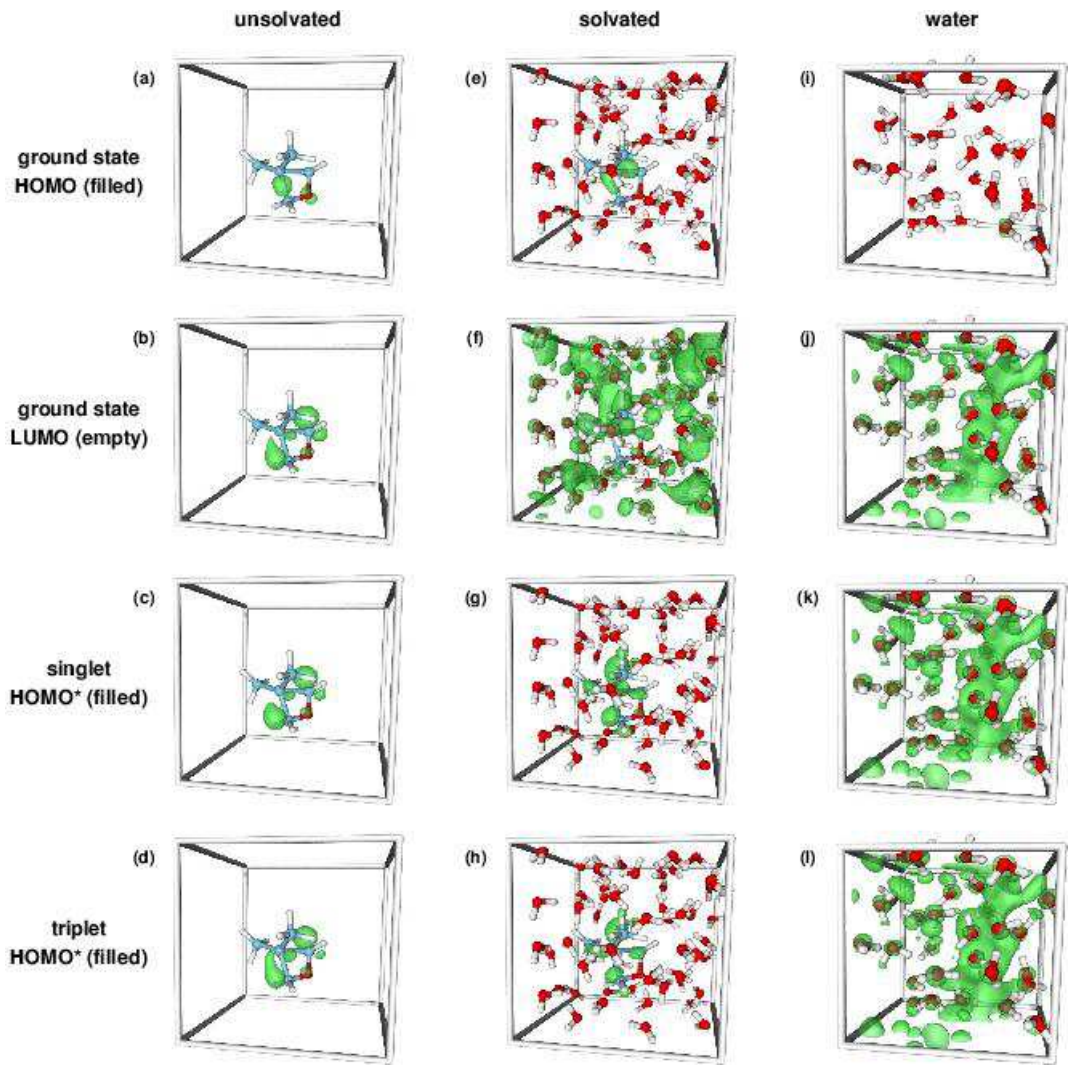}}
  \end{centering}
  \caption{
    The electronic structure of various Kohn-Sham eigenstates near the
    Fermi level for a specific configuration chosen at the 2.5~ps point
    from the FPMD simulation of 
    $\mbox{Si}_5\mbox{H}_{10}\mbox{O(II)}$ in water
    outlined in Fig.~\ref{fig.tempgap}. 
    The unsolvated systems (a)-(d) 
    contain equivalent solute structures to the solvated
    systems (e)-(h), but have all water molecules removed. 
    Also displayed (i)-(l) are states near the Fermi level for a configuration
    taken from an FPMD simulation of water at 300~K.
    All charge density 
    isosurfaces (green) contain 30\% of the total charge of the given state.
    Atom colors are as indicated in Fig.~\ref{fig.homolumo}. 
    }
  \label{fig.states}
\end{figure*}

We analyze the electronic structure of the solvated system
described in Sec.~\ref{Sec.gaps300K} in order to assess the validity
of our approach to the calculation of absorption gaps, and to
deduce why dielectric screening does not have a large impact on
the optical properties.
Figure~\ref{fig.states} illustrates
the charge density associated with several Kohn-Sham eigenstates
near the Fermi level, for
an isolated Si cluster (unsolvated), for the Si cluster surrounded by water
(solvated) and for a configuration taken from a FPMD
liquid water simulation at 300~K (water).

We examine first the states illustrated in Figure~\ref{fig.states}(e)-(h).
We notice that the HOMO of the solvated system [Figure~\ref{fig.states}(e)]
is localized on the Si
cluster solute. This localization of the HOMO 
is typical for each of the configurations examined in Fig.~\ref{fig.tempgap}. 
This is in contrast with, e.g., the acetone-water system
examined by Bernasconi 
{\it et al.}~\cite{BernasconiSprik2003} In that work, the HOMO varied in
its localization during the course of a FPMD simulation, at times
reminiscent of the HOMO of the isolated acetone system, and at others
more like the HOMO of water in the condensed phase.

The LUMO of the solvated system [Figure~\ref{fig.states}(f)] is entirely
delocalized over the simulation cell. However, when we alter the occupancy
of the Kohn-Sham eigenstates (as described in Sec.~\ref{Sec.AbinitioCalculations})
to produce the singlet or triplet first excited states, we notice that
the HOMO$^*$ -- the analogue of the LUMO in the occupied excited state --
is localized on the solute. In comparison with the localized 
states of the unsolvated Si cluster shown in Figure~\ref{fig.states}(a)-(d), 
only the LUMO of the solvated system differs significantly.

Our initial motivation for considering occupied excited state configurations
was to capture the screening response of the solvent to variations
in the solute charge density from its electronic ground state.
We find that the physically different charge distributions associated with the
LUMO and HOMO$^*$ of this solvated system are not significant for the
various estimates of the absorption gap. We see from 
Table~\ref{tab.gapcomparison} that the HOMO-LUMO estimate is
essentially the same as the $\Delta\mbox{SCF}$ singlet estimate.

\begin{table}[t]
\caption{Various DFT estimates of the absorption gap
         for the $\mbox{Si}_5\mbox{H}_{10}\mbox{O(II)}$ molecule:
         in its vacuum, zero temperature structure; 
         at 300~K {\it in vacuo}; and at 300~K in water.
         The finite temperature estimates of the gap use
         the molecular structure of
         the snapshot considered in Fig.~\ref{fig.states}.
         Also shown is the splitting between the singlet and triplet
         $\Delta\mbox{SCF}$ estimates of the gap.
         All energies given in eV.
        }
\begin{tabular*}{\columnwidth}{@{\extracolsep\fill}lccc}
  \hline \hline 
  Phase    & vacuum & vacuum & solvated \\
  Temp (K) & 0 & 300 & 300 \\
  \hline
  HOMO-LUMO                  & 4.5  & 3.7  & 3.9  \\
  $\Delta\mbox{SCF}$ singlet & 4.4  & 3.8  & 3.9  \\
  $\Delta\mbox{SCF}$ triplet & 4.2  & 3.6  & 3.8  \\
  singlet-triplet            & 0.19 & 0.18 & 0.17 \\
  \hline \hline
\end{tabular*}
\label{tab.gapcomparison}
\end{table}

Our use of the triplet excited state, for reasons of efficiency 
(Sec.~\ref{Sec.AbinitioCalculations}), is
justified by the close physical resemblance of the charge densities of
the HOMO$^*$ in each case.
The splitting between the singlet and triplet $\Delta\mbox{SCF}$ absorption
gaps can be associated with exchange effects local to the Si cluster,
since the relevant gap states are localized on the cluster and the
presence of water or the perturbation of the molecular structure has little
effect on this splitting, which remains $\sim0.2$~eV in all cases. 
 
However, the difference between the LUMO and HOMO$^*$ has important
consequences for other physical quantities, such as the oscillator strength
($f$) for this transition. We find that for the HOMO-LUMO transition, $f=0.05$,
while for the transition from LUMO$^*$ to HOMO$^*$, $f=0.60$
(refer to Fig.~\ref{fig.elevels}(b) for definitions of these states).
Using the LUMO$^*$ is justified given its almost complete (99.8\%)
overlap with the HOMO of the ground state.
This order of magnitude difference in $f$ will have important consequences for
calculated optical absorption. 

\begin{figure}[t]
  \begin{centering}
    \scalebox{0.6}{\includegraphics{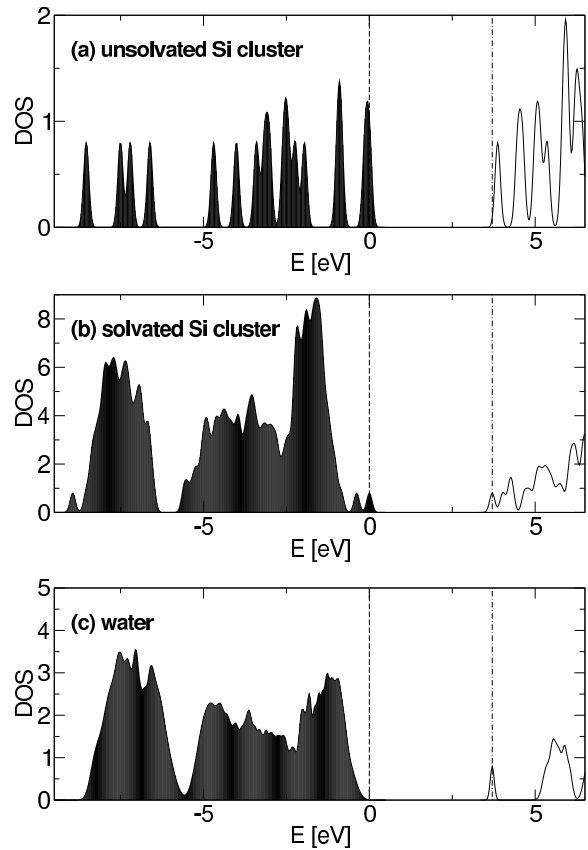}}
  \end{centering}
  \caption{
    The density of states (DOS) of (a) the unsolvated Si cluster of
    Fig.~\ref{fig.states}(a); 
    (b) the solvated Si cluster of Fig.~\ref{fig.states}(e); 
    and (c) that of liquid water.
    The DOS of water was calculated by thermal averaging over 20~ps
    of a FPMD simulation of 32 water molecules at ambient pressure and
    300~K. The average
    broadening of states for this simulation was 0.08~eV which was then
    used to broaden the discrete spectra of (a) and (b).
    The energy scale in (b) is shifted to place the HOMO at zero
    (vertical dashed line).
    The shift in energy applied to (a) places the HOMO at zero also,
    while the shift applied to (c) aligns the LUMO with the LUMO of (b)
    (vetical dashed-dotted line).
    The oxygen $s$-states have been omitted, but lie at $\sim-20\ \mbox{eV}$.
    }
  \label{fig.dos}
\end{figure}

The delocalization of the LUMO of the solvated system 
[Fig.~\ref{fig.tempgap}(f)] is reminiscent of the delocalized LUMO
observed in previous DFT calculations of the electronic structure
of liquid water.~\cite{LaasonenSprik1993,BoeroTerakura2001}
An example of this LUMO of water is provided in Fig.~\ref{fig.states}(j).
To interpret the origin of various states near the Fermi level
of the solvated system, 
we compare its density of states (DOS), in Fig.~\ref{fig.dos},
with that of the unsolvated system and that of pure liquid water.
The energy scale in this figure is shifted in such a way that 
the HOMO of the solvated system is at zero. 
Given that the HOMO and LUMO of the solvated system
are qualitatively similar to the HOMO of the isolated solute and 
the LUMO of pure water, 
respectively, we choose to align these pairs of states for the sake of
comparison.
It is worth noting that this seemingly arbitrary alignment is consistent for
other effective single-particle states. 
For example, the lowest valence $s$ states of the oxygen atoms
in each system are correctly aligned at around $\sim -20$~eV on this energy
scale (this part of the DOS has been omitted for clarity in Fig.~\ref{fig.dos}).
We also note that the isolated LUMO of liquid water, shown at $\sim$~2.5~eV
in Fig.~\ref{fig.dos}(c), is the subject of some controversy
and may not reflect a physically observable 
transition.~\cite{BernasconiSprik2003,BlumbergerBernasconi2004}

The DOS of the unsolvated Si cluster indicates another state
quite close in energy to the LUMO of the solvated system,
and, in fact, the LUMO+1 Kohn-Sham eigenstate of the solvated system 
resembles this unsolvated, localized, empty state.
This is also true of the HOMO$^*$ of the singlet and triplet configurations.
Therefore we conclude that occupying the first excited state leads to a
reordering of states near the Fermi level, such that the delocalized
LUMO is replaced by a localized HOMO$^*$. The 
preference for a localized HOMO$^*$ is due to the binding energy
associated with the underlying charge distribution, which contains
a ``hole'' (unoccupied LUMO$^*$) localized on the Si cluster.
To further illustrate this point, we see that for the states near the Fermi
level in liquid water [Fig.~\ref{fig.states}(i)-(l)] there is no
significant difference between the charge densities of the LUMO and
HOMO$^*$. There is no localized state nearby in energy
in the conduction band with which to swap, and also there is no localized
``hole'' to favor such a reordering of energy levels.

\section{Consequences for Experiment and Theory}
\label{Sec.Experiment}

There are several conclusions that experimentalists may 
draw from our computations.
With regard to reactivity, we see that for absorption processes, the
existence of the silanone group is very unlikely, given its low
activation energy for reaction with water. Note that the existence of
a photostable silanone in the presence of water, proposed by
Zhou and Head,~\cite{ZhouHead2000} is only relevant for emission processes.
Also, other work~\cite{ZhouBrus2003}
indicates that the significant red shift induced by 
this functional group may be reproduced by complete passivation with
other types of oxygen containing groups, such as hydroxyls and bridging
oxygens.

We expect that the large red shift caused by thermal strain in our
finite temperature simulations (Fig.~\ref{fig.tempgap}) may be
more pronounced for this small prototypical cluster than it would be
for a larger Si QD.
For larger Si QDs, with more rigid surface reconstructions,
we expect that finite temperature effects will play a smaller role
in their impact on the absorption gap. This is further reinforced by the
small temperature dependence of the lowest direct gap of bulk Si, 
which shifts to the red by only $\sim 0.05$~eV upon increasing the temperature 
from 0~K to 300~K.~\cite{AllenCardona1983}

The lack of any screening impact on the absorption of such a small
Si cluster, with a relatively large dipole moment indicates that
for larger Si QDs, this effect may also be negligible.
The likelihood of complete passivation of the surface of a large
Si QD with oxygen will significantly reduce the total dipole moment
of such particles. Furthermore, the preference for oxygen containing 
passivants which effectively preserve
the local tetrahedral symmetry around each Si atom guarantees that
the electronic states near the Fermi level are delocalized over the entire
QD. Therefore, screening impacts are reduced, given that they are more 
likely to be limited to the surface region of the QD, not extending far into
the core.
However, in larger Si QDs a smaller red shift
due to thermal strain may permit the observation of small 
solvent shifts due to screening.
This issue requires further investigation.

For theorists, it seems that simulating solvated Si QDs may not require
large FPMD simulations which include water. Given that the screening
impact is so small, there may be no advantage to including many solvation
shells of water molecules in a simulation. On the other hand, analysis of
ground state structures and active vibrational modes at a given finite
temperature may provide more useful information than large finite
temperature simulations and thermodynamic averaging.

\section{Conclusions}
\label{Sec.Conclusions}

This is the first theoretical analysis of the impact of aqueous solvation
on the optical properties of Si nanoparticles.
While there have been many studies on the impact of polar solvents
on the absorption properties of organic molecules, so far
no investigation has been reported for inorganic solutes such as QDs.
In our analysis we considered three factors:
chemical reactivity; thermal equilibration; and dielectric screening.
Regarding chemical reactivity, we find that 
the silanone functional group is extremely reactive in the presence of water
and is unlikely to exist in aqueous solvation.
We find that the bridging oxygen and hydroxide surface passivants are
much more stable and therefore more probable sources of red shifts in
the experimentally observed absorption spectra.
At 300~K, we find that thermal fluctuations induce strain
in the small silicon cluster examined here, and we estimate
that this strain leads to a systematic red shift of $\sim0.7$~eV.
Upon removing the screening impact of the surrounding water,
we determine no noticeable impact of solvent polarization or dispersion on
the optical absorption gap of this silicon cluster.
Given the thermal stability of larger QDs, we conclude that chemically
stable Si QDs will have robust optical properties in the presence of
water, further justifying their use as stable and efficient optical tags
in biological sensing applications.

\begin{acknowledgments}

We wish to thank F. Gygi, F. Reboredo and T. Ogitsu for 
stimulating discussions.
This work was performed under the auspices of the U.S. Department of Energy
at the University of California/Lawrence Livermore National Laboratory under
Contract No. W-7405-Eng-48.

\end{acknowledgments}

\bibliographystyle{achemso}

\providecommand{\refin}[1]{\\ \textbf{Referenced in:} #1}

\end{document}